\documentclass[prb,preprint]{revtex4-1} 
\usepackage{amsmath}  % needed for \tfrac, \bmatrix, etc.
\usepackage{amsfonts} % needed for bold Greek, Fraktur, and blackboard bold
\usepackage{graphicx} % needed for figures

\begin{document}

% Be sure to use the \title, \author, \affiliation, and \abstract macros
% to format your title page.  Don't use lower-level macros to  manually
% adjust the fonts and centering.

\title{The possibility of Scale Relativistic signatures in the Brownian motion of micro-spheres in optical traps}
% In a long title you can use \\ to force a line break at a certain location.

\author{Stephan LeBohec}
\email{lebohec@physics.utah.edu} % optional
\altaffiliation[permanent address: ]{201 JAMES FLETCHER BLDG. 115 S. 1400 E., Salt Lake City, UT 84112 , USA} % optional second address
% If there were a second author at the same address, we would put another 
% \author{} statement here.  Don't combine multiple authors in a single
% \author statement.
\affiliation{Department of Physics and Astronomy, University of Utah, Salt Lake City, UT 84112}
% Please provide a full mailing address here.

%\author{David P. Jackson}
%\email{ajp@dickinson.edu}
%\affiliation{Department of Physics, Dickinson College, Carlisle, PA 17013}

% See the REVTeX documentation for more examples of author and affiliation lists.

\date{\today}

\begin{abstract}
The development of a mechanics of non-differentiable paths \cite{mht1} suggested by Scale Relativity \cite{nottale1993,nottale2011} results in a foundation of Quantum Mechanics \cite{mht2,nottale2007} including Schr\"odinger's equation and all the other axioms under the assumption the path non-differentiability can be described as a Wiener process at the resolution-scale of observation. This naturally brings under question the possibility that the statistics of the dynamics of macroscopic systems fulfilling this hypothesis could fall under a {\it quantum-like} description with the Planck constant replaced with some other constant, possibly system specific, and corresponding to a diffusion coefficient. The observation of such a {\it quantum-like} dynamics would establish if the Scale Relativistic principle is implemented in macroscopic complex or chaotic systems. This would have major implications for the study of structure formation dynamics in various research fields. In this paper, I investigate the possibility for the detection of such an effect in the Brownian motion of a micro-sphere in an optical trap. I find that, if it exists, the observation of the transition to a {\it quantum-like} regime is within reach of modern experiments.  
\end{abstract}

\maketitle % title page 

\section{Introduction and background}
Scale Relativity \cite{nottale1993, nottale2011} is centered around the proposal to extend the relativity principle to changes of  resolution-scales. In other words, it revolves around the idea that, perhaps, reference frames are to be specified not only by their relative positions, orientations and motions but also by their relative resolution-scales. Then, the Scale Relativity principle is an extension of the usual relativity principle as it states that the laws of physics are the same in all reference frames independently from their relative positions, orientations, motions and resolution-scales \cite{nottale1993,nottale2011}.

When considering usual Newtonian mechanics, this does not make much of a difference: once the trajectories are observed or considered with a resolution-scale finer than their smallest convolutions, further refining the resolution-scale does not bring anything new. Newtonian mechanics implicitly  includes the assumption of differentiability for the trajectories followed by the system. Indeed, it is constructed from the consideration of infinitesimal displacements and time intervals.  For example, this is the case for the notion of velocity, while, in practice, a velocity is always obtained by taking the ratio between a finite displacement and a finite time interval, without ever actually taking the limit to infinitesimal intervals. Fundamentally, the infinitesimal limit does not correspond to a physical reality as the irregularities of a non ideal environment can always be expected to start affecting the trajectory when a fine enough resolution-scale is used. Even worse than this, if the resolution-scale is fine enough, it reveals the quantum domain. Then, by virtue of Heisenberg uncertainty relations, the outcomes of the velocity measurements  are increasingly affected as attempts are made to further approach the infinitesimal limit for displacements and time intervals. 

Indeed, R.P.Feynman and A.R.Hibbs\cite{feynmanhibbs1965}, in their 1965 path integral formulation of Quantum Mechanics, identified that the {\it "typical paths of a quantum-mechanical particle are highly irregular on a fine scale. Thus, although a mean velocity can be defined, no mean-square velocity exists at any point. In other words, the paths are non differentiable."} The length of such a quantum-mechanical path can therefore be shown to diverge like the inverse of the resolution-scale with which it is being inspected.  This was later described in terms of a fractal dimension \cite{mandelbrot1977}  $D_f=2$ for the quantum-mechanical path \cite{abbotwise1981,kroger}. This is also the fractal dimension of the path followed by a particle undergoing Brownian motion \cite{brown1866}. So it is clear that the assumed differentiability of mechanical paths in classical mechanics breaks down at least at small resolution-scales, giving way to resolution-scale dependence, characteristic of fractal objects. 

It then appears that the Scale Relativity program can be implemented in mechanics by considering non differentiable or fractal paths described at a resolution-scale which becomes an explicit parameter of the equations of dynamics. This will be outlined in Section \ref{SR_QM} and it is shown that, in the case of non differentiable paths resulting from a Wiener stochastic process, the equations of dynamics straightforwardly obtained are those of Quantum Mechanics. The derivation is based on the abandonment of the implicit hypothesis of differentiability but one particularly interesting aspect is that it does not depend on the statistics governing the stochastic process to be maintained all the way down to infinitesimal scales as is the case for standard Quantum Mechanics. The reasoning depends on the Wiener stochasticity only at the considered resolution-scale.  The important consequence of this is that it opens up the possibility for some  complex or chaotic macroscopic systems to display {\it quantum-like} features in their structure or dynamics. Indeed, some complex or chaotic systems considered over some range of resolution-scales, can be effectively described in terms of a Wiener process. These are situations in which the scale relativistic approach to Quantum Mechanics is applicable with of course the Planck constant $\hbar$ replaced by a system dependent diffusion constant characterizing the stochasticity. One could talk of stochastic \cite{Nelson1966} or macroscopic quantization. 

In fact, soon after the publication of Schr\"odinger's equation \cite{schrodinger1926}, it was noticed that the orbital parameters of the Solar system's planets and moons, which constitute a chaotic system \cite{laskar2009}, are arranged in a way very similar to the orbits in Bohr's model of the hydrogen atom \cite{caswell1929,malisoff1929,peniston1930}. This might be the first observation of the type of macroscopic {\it quantum-like} dynamics discussed here. These analyses were recently revisited in the Scale Relativity context \cite{nottale1997,hermann1998} and extended to other systems \cite{nottale2011,nottale2000,nottale1996}. Some of these results are impressive but it would be more convincing to observe similar dynamics in the laboratory under a controlled and reproducible environment.

In this article, I am trying to identify a way to investigate this type of {\it quantum-like} behavior in the laboratory. From the above comments, it is clear that the system must involve Brownian motion. In addition, an external field must be present to induce a Newtonian dynamic to confine the Brownian motion in such a way  {\it quantum-like} dynamics may establish itself. I propose that such a macroscopic {\it quantum-like} dynamics maybe observable in the motion of micro-sphere in an optical trap\cite{ashkin1986}. The surrounding fluid is then responsible for the Brownian motion while the optical trap effectively provides the external field.  In Section \ref{section_model}, I present the modeling of such a system in order to characterize the observability of {\it quantum-like} behavior neglecting dissipation whose effects are discussed in Section \ref{dissip}. In Section \ref{observation}, I review  the results of recent experiments in comparison with the expectations for the scale-relativistic effect. Finally, Section \ref{section_conclusion} summarizes the article and provides an outlook. Since Scale Relativity is not familiar, Section \ref{SR_QM} is used to give a very brief review of its connection with quantum mechanics.    

\section{Quantum Mechanics from the Scale Relativistic approach}\label{SR_QM}
There are two consequences to the abandonment of the differentiability hypothesis while considering the variation rate of a function $f(t)$. First the variation rate becomes doubled valued with possibly different variation rates  before and after any considered point. Second, at least when the function is non differentiable in a dense set of points, we cannot take the limit to infinitesimal time intervals anymore. In order to retain the use of differential calculus, we may define an explicitly resolution-scale dependent function $f(t,\delta t)$. For different values of $\delta t$, this function $f(t,\delta t)$ can be thought of as different approximations resulting from the sampling at different time intervals $\delta t$ of the underlying non-differentiable function $f(t)$. We then define a double-valued and  explicitly scale dependent finite differential: 
\begin{eqnarray*}
f'_+(t,\delta t)&=&\frac{f(t+\delta t,\delta t)-f(t,\delta t)}{\delta t}{~~~~\delta t>0};\\
f'_-(t,\delta t)&=&\frac{f(t+\delta t,\delta t)-f(t,\delta t)}{\delta t}{~~~~\delta t<0}.
\end{eqnarray*}

We may apply this to the description of a displacement along a non-differentiable path: 
\begin{eqnarray*}\label{repres}
d{\bf x}_+&=&{\bf v}_+dt+d{\bf b}_+{~~~~d t>0} \\
d{\bf x}_-&=&{\bf v}_-dt+d{\bf b}_-{~~~~d t<0}
\end{eqnarray*}
The first term proceeds from a {\it usual velocity} ${\bf v}_\pm$,while, $d{\bf b}_\pm$ represents a possibly stochastic residual $\langle d{\bf b}_\pm\rangle=0$ accounting for the details of the path at a finer resolution-scale.  This two term description formalizes the common practice of disregarding or smoothing-out the details smaller that some resolution-scale. This is fully implemented by taking the expectation values of the time-differentials {\it after} and {\it before} the considered instant:
$$
\frac{d_+}{dt}{\bf x}={\bf v}_+ +\langle\frac{d{\bf b}_+}{dt}\rangle={\bf v}_+{\rm ~~and~~} \frac{d_-}{dt}{\bf x}={\bf v}_-+\langle\frac{d{\bf b}_-}{dt}\rangle={\bf v}_-
$$
and it is convenient to combine them linearly into a single {\it complex time-differential operator}\,\cite{nottale1993}: 
\begin{eqnarray*}\label{complex_diff_op}
\frac{\hat d}{dt}=\frac{1}{2}\left(\frac{d_+}{dt}+\frac{d_-}{dt}\right)-\frac{i}{2}\left(\frac{d_+}{dt}-\frac{d_-}{dt}\right).
\end{eqnarray*}
This operator can be used to define a complex velocity:
$$
\mathcal V=\frac{\hat d}{dt}{\bf x}=\frac{{\bf v}_++{\bf v}_-}{2}-i\frac{{\bf v}_+-{\bf v}_-}{2}={\bf V}-i{\bf U}
$$
in which the real part $\bf V$ is the {\it classical velocity} while the imaginary part $\bf U$, the {\it kink velocity} reveals the non differentiable nature of the path at the considered resolution-scale.

We may then write the complex time-differential of a regular and indefinitely differentiable  field $h({\bf x}(t, dt),t)$ along a resolution-scale dependent path ${\bf x}(t,dt)$.  In doing so, we make the specific choice for the residual stochastic process $d{\bf b}_\pm$ to be a Wiener process with  $\langle d{\bf b}_\pm\rangle=0$, $\langle db_{i+}\cdot db_{i-}\rangle=0$,  and $\langle db_{i+}\cdot db_{j+}\rangle=\langle db_{i-}\cdot db_{j-}\rangle=2\mathcal D\delta_{i,j}dt$ with $\mathcal D$ a diffusion coefficient. Keeping only the Taylor expansion terms of $\frac{\hat dh}{dt}$ that do not vanish with $dt$, the complex time-differential takes the form\cite{nottale1993,nottale2011,mht1}: 
\begin{eqnarray}\label{covardiff}
\frac{\hat d}{dt}=\frac{\partial }{\partial t}+\mathcal V\cdot \nabla -i \mathcal D\Delta 
\end{eqnarray}

If the scale relativistic principle applies, the stochastic system must be characterized by a complex Lagrange function $\mathcal L({\bf x},\mathcal V,t)$ and enforcing the stationarity of the  action $\mathcal S=\int_{t_1}^{t_2}\mathcal L({\bf x},\mathcal V,t)dt$ (while carefully keeping track of the changes in the Leibniz product rule resulting from the higher order differential term which appeared in $\frac {\hat d} {dt}$) results in the usual Euler-Lagrange equation with the complex time-differential operator and complex velocity replacing the usual time derivative and usual velocity: 
$$
\nabla_{\bf x}\, \mathcal L-\frac{\hat d}{dt}\nabla_{\mathcal V} \,\mathcal L=0
$$
In particular, with  $\mathcal L=\frac{1}{2}m\mathcal V^2-\Phi$ where $\Phi$ is a purely real potential energy, we recover a generalized form of Newton's relation of dynamics
\begin{eqnarray}\label{newton}
m\frac{\hat d}{dt}\mathcal V=-\nabla\Phi.
\end{eqnarray}

 In the stationary case, $\langle{\bf V}\rangle=0$, this generalized relation of dynamics can be turned into a Langevin equation which has been integrated numerically \cite{mclendon, hermann1997,alrashid2011,mht1} for various systems. In each case, the statistics of the system configuration parameters are found to follow the magnitude squared of the solutions of the time independent Schr\"odinger equation describing the same system with however Planck's constant replaced by $2m\mathcal D$. This is particularly striking as standard Quantum Mechanics has not been invoked in the establishment of the above generalized relation of dynamics (Equation \ref{newton}). 

The connection with standard Quantum Mechanics can be made explicit by introducing the  function $\psi=\psi_0 e^{i\mathcal S/\mathcal S_0}$ where $\mathcal S_0$ and $\psi_0$ are introduced for dimensional reasons. The complex velocity can then be expressed canonically as  $\mathcal V=-i\frac{\mathcal S_0}{m}\nabla \ln \left(\psi/\psi_0\right)$  which leads to rewriting the generalized Newton's relation of dynamics as a Schr\"odinger equation in which $\hbar$ is replaced with $\mathcal S_0=2m\mathcal D$ \cite{nottale1993, nottale2011}:
$$ 2im\mathcal D{{\partial\psi}\over{\partial t}} =-2m\mathcal D^2 \Delta\psi+\Phi\psi$$

I only provide an outline of this development here. Details as well as interpretive discussions of the other postulates of quantum mechanics can be found elsewhere \cite{nottale2007,nottale1993,nottale2011, mht1, mht2}. What is important is that this does not result from anything else than the relaxation of the mechanical path differentiability hypothesis under the restriction to Wiener processes for the stochastic component. Also it does not require the stochasticity to be preserved at resolution-scales different from  those considered in the observation or description of the system. The enforcement of the principle of Scale Relativity for mechanical path with a Wiener stochastic component results in this generalized Schr\"odinger equation. Consequently, it becomes justified to expect systems effectively evolving along this type of paths to display {\it quantum-like} dynamics.  This is precisely what I take under consideration in the next section with the motion of a micro-sphere in an optical trap. 

\section{Motion of micro-spheres in a harmonic traps}\label{section_model}
Consider a sphere of radius $R$ and mass density $\rho_I$ in a fluid of mass density $\rho_E$. The inertial mass $m$ of the sphere includes an added mass term 
representing the inertia of the fluid the sphere displaces in the course of its motion $m={{4}\over{3}}\pi R^3(\rho_I+\rho_E/2)$. The micro-sphere is in a trap which I assume to be harmonic $\Phi(r)=\frac{1}{2}m\omega^2 r^2$ with $\omega$ the proper frequency. In principle, the distribution of the position of the micro-sphere in the harmonic potential follows Boltzmann's statistics $e^{-\Phi(r)/k_BT}$ and for the harmonic potential we can expect a Gauss distribution with the standard deviation:
\begin{equation}
\label{classicboltz}
\sigma_B=\sqrt{{k_BT}\over{m\omega^2}}
%=\sqrt{{3k_BT}\over{4\pi R^3 (\rho_i+\rho_e/2)\omega^2}}
\end{equation}
%={{m\omega^2}\over{6\pi\eta R}}r
This Boltzmann distribution does not depend on any property of the fluid other than its temperature $T$ and its mass density, which enters in the micro-sphere's effective mass $m$. It is not achieved instantaneously. The harmonic force acting on the micro-sphere may result in a mobility limited radial drift with a speed $\langle v_r\rangle=-\mu m \omega^2 r$  where the mobility is given by Stokes law as ${\mu}=\frac{1}{6\pi\eta R}$ where $\eta$ represents the fluid viscosity. It is convenient to  define a relaxation time constant $\tau_r={{3\pi\eta R}\over{m\omega^2}}$. It depends on the viscosity $\eta$ but not on the temperature.  After a time $t\gg \tau_r$, the system reaches a statistically stationary state, independently from the initial configuration. This distribution results from a Langevin dynamics equation \cite{clercx1992}
$$
m\ddot{\bf r}=-m\omega^2{\bf r}-6\pi\eta R\dot{\bf r} +\cdots+{\bf F}_{Therm.}
$$
where ${\bf F}_{Therm.}$ represents the stochastic thermal force and I omit additional hydrodynamics forces which will be mentioned in Section \ref{dissip}.

The previous section suggests that, instead, the state of the micro-sphere should be described in terms of {\it quantum-like} wave functions $\psi({\bf r},t)$, solutions of a Schr\"odinger equation with the reduced Planck constant replaced with $2m\mathcal D$,
$$ i\mathcal D{{\partial\psi}\over{\partial t}}+\mathcal D^2 \Delta\psi =\frac{1}{2}\omega^2 r^2\psi+\cdots$$
where I omit all hydrodynamics forces including Stoke's drag, whose effect will be discussed in Section \ref{dissip}.

\begin{figure}
\centering
\includegraphics[width=4in]{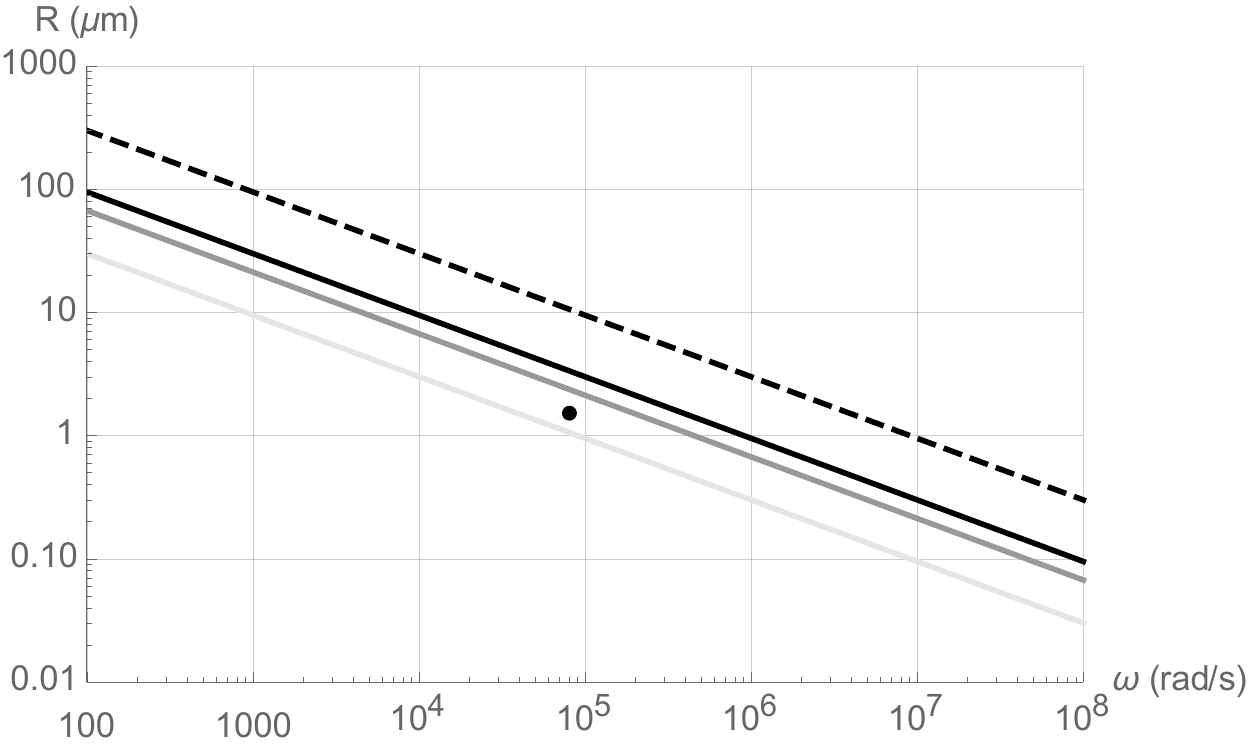}
\caption{The silica ($\rho_I=2.2\,\rm g\,cm^{-3}$) micro-sphere radius $R$ is represented as a function of the harmonic oscillator proper frequency  $\omega$ with their relation constrained by the requirement that $\bar n=0.1$ (dashed line) $\bar n=1$ (solid black line), $\bar n=2$ (grey line) and $\bar n=10$ (light grey line). The fluid is assumed to be water with a density of $1\,\rm g\,cm^{-3}$ and a $10^\circ\rm C$ viscosity $\eta=10^{-3} \,\rm Pa\,s$. The point corresponds to the system of a silica micro-sphere of diameter $3.06\,\rm\mu m$ in an optical trap with strength $188\,\rm\mu N \,m^{-1}$ in water used for a test of the Maxwell-Boltzmann distributions by J.Mo et al.\cite{mo2015}}
\label{fig1}
\end{figure}

The eigen-energies of the time independent version of this Schr\"odinger equation are  $E_n=(n+\frac{1}{2})2m\mathcal D \omega$ in the one dimensional case. Since the micro-sphere is in a bath at a given temperature $T$, its {\it quantum-like} state is a Boltzman factor weighted mixture of Hermite functions. The average {\it occupation number} of each state is proportional to $e^{-(n+1/2)/\bar n}$, with $\bar n=\frac{k_B T}{2m\mathcal D \omega}$,  where $k_B$ is Boltzmann's constant. As the micro-sphere is subject to Brownian motion,   the diffusion coefficient is given by Einstein's kinetic theory relation \cite{einstein1905,smoluchowski1906} as $\mathcal D=\mu k_B T$. So we have $\mathcal D={{ k_B T}\over{6\pi\eta R}}$ and $\bar n=\tau_r \omega$.  While this description is based on the application of Boltzmann's statistics to {\it quantum-like} states, if it has any validity, it should result in some departure from the Boltzmann statistics for  the position and motion of the micro sphere, at least in the {\it quatum-like regime} where $\bar n\ll 1$.  Figure \ref{fig1} shows the relations between the radius $R$ of a micro-sphere made of amorphous silica and the proper frequency of the trap $\omega$ for various values of $\bar n$ in water at a temperature of $10^\circ\rm C$. While $\bar n$ does not depend explicitly on temperature, it depends on the fluid viscosity $\eta$, which depends on temperature. For each micro-sphere radius, there is an optical trap strength above which {\it quantum-like} behavior should appear. For reference, the dot represents conditions used in observations by J.\,Mo et al.\cite{mo2015} and corresponding to $\bar n\approx 6.7 $. The same group performed other measurements under conditions corresponding to even smaller values of $\bar n$ which are more favorable to the observation of a scale relativistic signature as we are about to discuss.  

If the scale relativistic {\it quantum-like} description of the state of the micro-sphere undergoing Brownian motion in an optical trap is valid, the statistical distribution of the positions of the micro-sphere should correspond to the Boltzmann coefficient weighted series of the magnitude squared stationary {\it quantum-like} wave functions. In the present case of a harmonic well, these wave functions are Hermite functions.  In the first volume of statistical physics in the course of theoretical physics by L.D.~Landau and E.M.~Lifshitz  \cite{landaustatmech} , this distribution is shown to be gaussian. Replacing $\hbar$ with $2m\mathcal D$ and making use of  $\bar n$ in the expression of the standard deviation, we find
%$$
%\sigma_Q=\left[{\tanh(\frac{2R^2\omega\left(\rho_i+\rho_e/2\right)}{9\eta})\frac{6\pi\eta\omega R}{k_B T}}\right]^{-1/2}.
%$$
%$$
%\sigma_Q=\left[{\tanh(\frac{4\pi R^3\omega\left(\rho_i+\rho_e/2\right)}{18\pi\eta R})\frac{6\pi\eta\omega R}{k_B T}}%\right]^{-1/2}.
%$$
%\begin{equation}
%\label{quantumlike}
%\sigma_Q=\left[{\tanh\left(\frac{\omega m}{6\pi\eta R}\right)\frac{6\pi\eta\omega R}{k_B T}}\right]^{-1/2}.
%\end{equation}
%\begin{equation}
%\label{quantumlike}
%\sigma_Q=\left[{\tanh\left(\frac{1}{2\bar n}\right)\frac{2m\omega^2}{k_B T}\bar n}\right]^{-1/2}.
%\end{equation}

\begin{equation}
\label{quantumlike}
\sigma_Q=\sigma_B\left[{2\bar n\,\tanh\left(\frac{1}{2\bar n}\right)}\right]^{-1/2}.
\end{equation}

\begin{figure}
\centering
\includegraphics[width=4in]{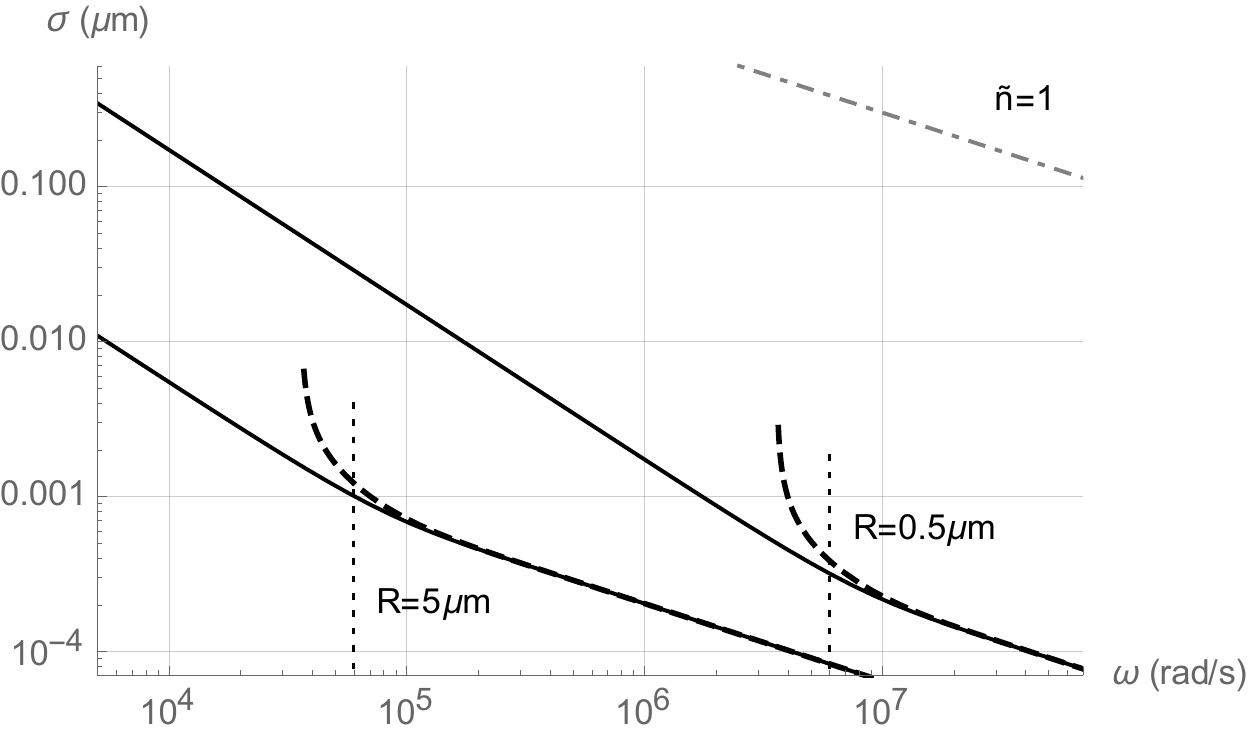}
\caption{The calculated standard deviation of the position of a micro-sphere of radius  $R=5\mu m$ and  $R=0.5\mu m$ is shown as a function of the harmonic trap proper frequency $\omega$. Conditions are the same as in Figure \ref{fig1}. The vertical dotted lines mark the respective values of $\omega_0$ around which the transition from the standard Boltzmann statistics to quantum-like statistics may occur. The transition is revealed by the change of slope of the curves in this Log-Log graph, indicating a change from a $\frac{1}{\omega}$ dependence for $\omega\ll\omega_0$ to a $\frac{1}{\sqrt\omega}$ dependence for $\omega\gg\omega_0$. The dashed curves represent the calculated standard deviation when viscous dissipation is taken into account as discussed in Section \ref{dissip}. The grey dot-dashed line at the top of the figure is shown for reference. It indicates the radius of the micro-sphere such that $\bar n=1$. }
\label{fig2}
\end{figure}

In the limit of a weak trap for which $\bar n \gg 1$, we see that $\sigma_Q\to\sigma_B$. In this regime, {\it quantum-like} effects are not expected and the present theory reproduces the classical result. 
However, in the limit of a strong trap for which $\bar n\ll 1$, we may expect the {\it quantum-like} ground state to dominate the statistics with a different dependence on the conditions:  
%$$\sigma_{Q_0}=\sqrt{{{\mathcal D}\over{\omega}}}=\sqrt{{{k_B T}\over{6\pi\eta R\omega}}}$$
%$$\lim_{\bar n\to 0}\sigma_Q=\frac{\sigma_B}{\sqrt {2\bar n}}=\sqrt{{{k_B T}\over{6\pi\eta R\omega}}}$$
$$\lim_{\bar n\to 0}\sigma_Q=\sqrt{{{k_B T}\over{6\pi\eta R\omega}}}$$
Starting from the classical regime and keeping all things equal, as the strength of the trap is increased, the standard deviation of the micro-sphere position decreases as $\frac{1}{\omega}$. The observation of a transition from a $\frac{1}{\omega}$ to a $\frac{1}{\sqrt{\omega}}$ dependence of the standard deviation of the position micro-sphere in the vicinity of $\omega_0=\frac{3\pi\eta R}{m}$ would be a signature of the actual existence of a scale relativistic or {\it quantum-like} regime.  This possible transition is illustrated by Figure \ref{fig2} for two different silica micro-sphere radii in water. Figure \ref{fig2} also shows that the observation of the transition to the {\it quantum-like}  regime with predominance of the ground state requires the ability to monitor the motion of the micro-sphere with a resolution that is more that three orders of magnitude smaller than the size of the microsphere. Amazingly, observations have already been performed in the domain of interest. However, before reviewing them, effects neglected in the above discussion should be discussed as they may affect the observation of the signature transition. 

\section{Effects of dissipation and hydrodynamics}\label{dissip}
The expression of $\sigma_Q$ (Equation \ref{quantumlike}) corresponds to an ideal harmonic oscillator while, in fact, the micro-sphere also is under the influence of a dissipative drag force.  Quantizing the  so-called Bateman dual system of damped and amplified oscillators, M.~Blasone and P.~Jizba showed \cite{blasone2002} that, in first approximation, the quantum ground state of the under-damped harmonic oscillator to have the same form as the quantum ground state of the simple harmonic oscillator with the proper frequency of the under-damped oscillator $\tilde\omega=\omega\sqrt{1-(2m\omega\mu)^{-2}}$ used in the place of the proper undamped frequency $\omega$. With the Stokes' drag as the dissipative force, we have 
%$\tilde\omega=\sqrt{\omega_0^2-\frac{\gamma^2}{4m^2}}$. \\
%$\tilde\omega=\omega\sqrt{1-\frac{1}{4m^2\omega^2\mu^2}}$. \\
 $\tilde\omega=\omega\sqrt{1-\bar n^2}$. 
We see that the boundary of the domain within which the oscillator is under-damped corresponds to $\bar n=1$. This is precisely where we expected the transition to the {\it quantum-like} regime. With this damped oscillator frequency, we may define the damped equivalent of $\bar n$ which we write $\tilde{\bar n}=\frac{k_B T}{2m\mathcal D \tilde\omega}$ and we find $\tilde{\bar n}=\frac{\bar n}{\sqrt{1-\bar n^2}}$.  For $\bar n\ll1$, that is far in the under-damped oscillation regime, we have $\tilde{\bar n}\approx\bar n$ and nothing is changed. When we just have $\bar n< 1$, we see that  $\tilde{\bar n}\ge\bar n$ so the dissipative effects tend to move the system away from the {\it quantum-like} regime, particularly when $\bar n$ is close to unity.  At the same time, as the dissipative term results in a smaller effective harmonic force,  the resulting standard deviation $\tilde\sigma_Q$ may be larger than $\sigma_Q$. As an indication of this, $\sigma_Q(\tilde\omega)$ is also shown on Figure \ref{fig2}. However, if the dissipation is taken into account, for $n>1$ the above theory is simply not applicable. Nevertheless, this does not say anything about the Scale Relativity principle being implemented in nature or not. When the drag force is important, both the motion of the micro-sphere and the motion surrounding fluid should be described using {\it quantum-like mechanics}. It remains that if the scale relativistic principle is implemented in nature, the above described transition between the classical domain and {\it quantum-like} domain, with $\bar n\ll 1$, where dissipative effects are negligible, should still be observable.

Other effects can complicate the comparison between the classical and {\it quantum-like} regimes. When the velocity of an object immersed in a fluid changes, the boundary layer adjusts with a time delay. Correspondingly, such changes in the vorticity of the fluid affects the force acting on the object at a later time in a way depending on the object motion history, this is the Boussinesq-Basset force \cite{feng2012}. I find that, in the quantum regime $\bar n\ll 1$, it only constitute a small correction to Stokes' drag, which we just discussed.  Further more, when considering time scales shorter than the size of the object divided by the speed of sound, in principle, the finite compressibility of the fluid should also be taken into account \cite{zwanzig1975}. 

These effects can be regarded as departures from a purely Brownian motion, corresponding to a fractal dimension $D_F=2$, the only one for which the Schr\"odinger equation is strictly valid. In principle, to be included in a Scale Relativity description, these effects would require a generalized Schr\"odinger equation, not restricted to strictly Markovian processes. Here I simply neglect to include these effects and we should expect that the above described transition do not necessarily follow the  black curves of Figure \ref{fig2} with precision but it remains that a change of behavior in the regime $\bar n\ll 1$ is to be expected if the Scale Relativity principle is implemented. 

\section{Observations}\label{observation}
Several experiments have been performed to track the motion of a micro-sphere in an optical trap \cite{ashkin1986} in a fluid. In these experiments, the light from one of the lasers forming the optical trap is collimated after scattering off the micro-sphere and directed toward a system of balanced detectors. The resulting signal  depends on the position of the micro-sphere in the trap. It is digitized and recorded at a high rate for extended periods of time \cite{li2010,mo2015}. The statistics of the data is then analyzed by comparison to the Ornstein-Uhlenbeck model \cite{uhlenbeck1930} with the addition of the effective mass and the contribution of Basset force in liquids\cite{kheifets2014}.  For measurements in a gas the data statistics is compared to the solution of the Langevin equation in an underdamped harmonic trap as given by Wang and Uhlenbeck \cite{wang1945}. In both cases the equipartition theorem is satisfied for instantaneous mean squared velocity and position in the harmonic well  provided the effective mass $m$ of the the micro-sphere is used. 

In order to test for  a {\it quantum-like} behavior as suggested by the scale relativistic approach outlined above, we may concentrate on the distribution of the position of the micro-sphere. Table \ref{tablemo} summarizes the experimental parameters presented in the articles by J.~Mo et al. \cite{mo2015} and T.~Li et al. \cite{li2010} and includes the calculated value of $\bar n$ and $\frac{\sigma_Q}{\sigma_B}-1$ for each. Both measurements with silica micro-sphere in liquids correspond to $\bar n>1$ and do not provide much of an opportunity to discriminate between $\sigma_B$ and $\sigma_Q$. However, the two other measurements provide $\bar n<1$ with appreciable differences of 6.4\% and 20\% respectively between  $\sigma_B$ and $\sigma_Q$, which could lead to a clear detection or exclusion of the {\it quantum-like} effect. 

%\begin{table}[htdp]
\begin{table}
\caption{ \label{tablemo}In the following table, the values of $R$, $m$, and $\omega$ are from the measurements presented in the articles by J.~Mo et al. \cite{mo2015} and T.~Li et al. \cite{li2010}. These quantities are used to calculate $\bar n$ as in Section \ref{SR_QM} as well as the relative difference between $\sigma_Q$ and $\sigma_B$. 
Calculations were done using the following values specified by J.~Mo et al. \cite{mo2015} for measurements in liquid phases:  
$\rho_{\rm Silica}=2.0\,\rm g\cdot cm^{-3}$, 
$\rho_{\rm BaTiO_3}=4.2\,\rm g\cdot cm^{-3}$, 
$\rho_{H_2O}=0.998\,\rm g\cdot cm^{-3}$, 
$\rho_{Acetone}=0.789\,\rm g\cdot cm^{-3}$,  
$\eta_{H_2O}=9.55\times 10^{-4}\,\rm Pa\,s$, and
$\eta_{Acetone}=3.17\times 10^{-4}\,\rm Pa\,s$ and we used $\omega=\sqrt{k/m}$ where $k$ is the trapping strength given in $\mu N/m$. 
For the measurement in the air, we used 
$\rho_{\rm Silica}=2.2\,\rm g\cdot cm^{-3}$, 
$\rho_{Air}\approx 0$  and 
$\eta_{Air}=1.81\times 10^{-5}\,\rm Pa\,s$}
\begin{center}
\label{solartable}
%\resizebox{\textwidth}{!}
{
\begin{tabular}{|c|c|c|c|c|}
\hline
Systems  &    ${\rm Silica~in}\atop{\rm Water}$  &  ${\rm Silica~in}\atop{\rm Acetone}$  & ${\rm BaTiO_3~in}\atop{\rm Acetone}$ & ${\rm Silica~in}\atop{\rm air}$\\
\hline
$R$ ($\rm \mu m$)&                           1.53 &  1.99 &  2.68 & 1.50\\
$m$ ($\rm 10^{-14}\,kg$)   &              2.25 &   5.30 &  30.7 & 3.11 \\
$\omega$ ($\rm 10^4\,s^{-1}$)   &      9.14 &  3.07 &  3.34 & 1.99\\
\hline
$\bar n$   &                                         6.69 &  3.65 &  0.78 & 0.41\\
$\frac{\sigma_Q}{\sigma_B}-1$   &      $9.3\times 10^{-4}$   & $3.1\times 10^{-3}$  & $6.4\times 10^{-2}$ & 0.20\\
\hline
\end{tabular}}
\end{center}
\label{default}
\end{table}

The papers presenting the measurements do not report any incompatibility between the above mentioned models to the data statistics. However the physical parameters that are kept free in the fitting procedures are the strength of the harmonic well $\omega$, the micro-sphere radius $R$ in the case of the measurements in liquids \cite{mo2015} and for the measurements in air \cite{li2010} the free parameters are the strength of the harmonic well $\omega$ and the momentum relaxation time $\tau_p=\frac{m}{6\pi\eta R}$. We have seen that $\sigma_B$ is inversely proportional to $\omega$. A difference of a few percents in the spread of the micro-sphere position distribution could be absorbed by a few percent change in the well strength. Possibly more importantly, the calibration factor $K$ used to convert the balanced detector output signal to an actual position is also kept free, at least in the analysis of the measurements in liquids \cite{mo2015} while no details are given for the calibration of the position detector used in the measurements in the air. The factor $K$ is a scaling factor for the micro-sphere position distribution and it could as well absorb a few percent difference. As a consequence the difference between $\sigma_Q$ and $\sigma_B$ might have gone unnoticed as it could have been absorbed in fit parameters.  

A search for a transition around $\bar n\sim 1$ could start in the regime $\bar n\gg 1$ where all the calibration could be performed. Then, the strength of the harmonic well could be increased progressively to bring the system to the regime $\bar n\ll 1$. This supposes the strength of the well can be adjusted without affecting the position detector calibration. It also requires some mean of independently calibrating the strength of the well over at least an order of magnitude. 

\section{Summary}\label{section_conclusion}

The Scale Relativity principle extends the usual relativity principle to include changes of resolution-scale and leads to considering non-differentiable mechanical paths. In order to retain differential calculus as a tool, these paths maybe considered at a set resolution scale which becomes an explicit parameter. Then the development of a mechanics for paths with their non-differentiable component resulting from a Wiener process leads to a generalized newtonian relation of dynamics, which cas be rewritten as a Schr\"odinger equation in which $\hbar$ is replaced by $2m\mathcal D$ where $m$ is the mass of the micro-sphere and $\mathcal D$ is a diffusion constant. If the Scale Relativity principle is implemented in nature, chaotic or complex systems whose evolution can effectively be described as a Wiener stochastic process should display a dynamic structure in accordance to this {\it emergent quantum-like mechanics}. 
 
In particular the Brownian motion of a micro-sphere in an optical trap could reveal such a behavior. In Section \ref{section_model}, we have seen that the Boltzmann statistics occupation of the {\it quantum-like} levels of the harmonic oscillator reproduces the classical position probability density in the limit $\frac{3\pi\eta R}{m\omega}\gg1$ with a spread  scaling with $\frac 1 \omega$. However, when $\frac{3\pi\eta R}{m\omega}\ll1$, the statistics of the {\it quantum-like} ground state may dominate and the spread of the position probability density then scales with $\frac {1}{ \sqrt{\omega}}$. In Section \ref{dissip} we have seen that the dissipative drag force and other effects may affect the details of the transition between the two regimes but are not expected to prevent distinguishing them observationally.  The observation of the change of behavior between the two regimes would be indicative that the Scale Relativity principle is actually  implemented in nature. 

In Section \ref{observation} we have seen that modern experiments \cite{mo2015,li2010} used to record  the motion of a micro-sphere in an optical trap have a sensitivity that allows probing the  domain where the transition between classical and {quantum-like} regimes may occur. While these experiments did not report any departure from the classically expected behavior, I argued that the relatively small relative difference in the spread of the micro-sphere position probability density could have been absorbed in the calibration factor of the position measurement and in the strength of the optical well, which are both fit parameters in the reported data analysis. 

Observational evidences for the type of {\it macroscopic stochastic quantization} discussed in this paper would have major implications for many research fields such as astrophysical structure formation, geology, meteorology, biology, ecology, sociology or economy as the assumed underlying dynamics may in some cases have to be revisited.  

\begin{acknowledgments}
The author is grateful to Mark Raizen for his clarifications and encouragements. The author also acknowledges  Laurent Nottale, Patrick Fleury, Eugene Mishchenko and Janvida Rou for their very helpful comments and suggestions.

\end{acknowledgments}

\end{document}